\documentclass[sigconf]{acmart}
\usepackage{tabularx}
\usepackage{multirow}
\usepackage{csquotes}
\usepackage[font=small,labelfont=bf]{caption}

\usepackage{booktabs} 

\setcopyright{rightsretained}

\acmConference[ESEC/FSE 2018]{}{November 2018}{Florida, US} 
\acmYear{2018}
\copyrightyear{2018}

\begin{document}
\title{Estimating defectiveness of source code: A predictive model using GitHub content}

\author{Ritu Kapur}
\orcid{0000-0001-7112-0630}
\affiliation{%
  \institution{Indian Institute of Technology}
  \city{Ropar} 
  \state{Punjab} 
  \postcode{140001}
}
\email{ritu.kapur@iitrpr.ac.in}

\author{Balwinder Sodhi}
\orcid{}
\affiliation{%
  \institution{Indian Institute of Technology}
  \city{Ropar} 
  \state{Punjab} 
  \postcode{140001}
}
\email{sodhi@iitrpr.ac.in}

\renewcommand{\shortauthors}{Ritu Kapur, Balwinder Sodhi}

\begin{abstract}
Two key contributions presented in this paper are: \textbf{i)} A method for building a dataset containing source code features extracted from source files taken from Open Source Software (OSS) and associated bug reports, \textbf{ii)} A predictive model for estimating defectiveness of a given source code. These artifacts can be useful for building tools and techniques pertaining to several automated software engineering areas such as bug localization, code review, and recommendation and program repair.

In order to achieve our goal, we first extract coding style information (e.g. related to programming language constructs used in the source code) for source code files present on GitHub. Then the information available in bug reports (if any) associated with these source code files are extracted. Thus fetched un(/ semi)-structured information is then transformed into a structured knowledge base. We considered more than $30400$ source code files from $20$ different GitHub repositories with about $14950$ associated bug reports across $4$ bug tracking portals. The source code files considered are written in four programming languages (viz., C, C++, Java, and Python) and belong to different types of applications. 

A machine learning (ML) model for estimating the defectiveness of a given input source code is then trained using the knowledge base. In order to pick the best ML model, we evaluated 8 different ML algorithms such as Random Forest, K Nearest Neighbor and SVM with around $50$ parameter configurations to compare their performance on our tasks. One of our findings shows that best K-fold (with k=5) cross-validation results are obtained with the NuSVM technique that gives a mean F1 score of $0.914$.
\end{abstract}

%
%


\maketitle

\section{Introduction}
Maintenance of software, particularly identifying and fixing the defects, contributes significantly towards the overall cost \cite{bennett2000software, lientz1980software} of developing and deploying a software application. It has been observed \cite{deissenboeck2006concise,allamanis2014learning} that coding practices adopted by the programmers greatly influence the quality of software. We believe that an in-depth comprehension of how the \textit{programming styles} influence \textit{defectiveness} of software can help in avoiding software defects. Availability of a large volume of source code from Open Source Software (OSS) and its associated defect reports present a ripe opportunity to apply Artificial Intelligence (AI) techniques to build a system for estimating the presence of defects in a given source code. 

One of our important goals is to be able to determine the \textit{defectiveness} of a source code by considering the \textit{programming style} used in the code. In other words, a broader question (and aim for us) is to check if there exist any correlation between the programming style and the quality of a software. We believe that \textit{Among two software applications developed using similar processes, the one having fewer reported defects is more likely to be of better quality than the other which has more defects reported against it.} The \textit{defectiveness} for us includes two things: \textbf{i)} likelihood of finding defects of different types and severity etc. expressed on the scale \{\texttt{Likely-to-be-defective, Unpredictable}\}, and \textbf{ii)} the properties of potential defects that can be present in a given source file. Such properties include: type, severity, project phase in which reported etc.

Broadly speaking, our system can answer two questions for an input source file: \textit{whether it is likely to be defective}, and if yes then \textit{what are the properties of likely defects}. We use a binary classification methodology while answering these questions.\\ 

\textbf{Leveraging information available on OSS projects}\\
OSS repositories like GitHub\footnote{https://github.com/}, SourceForge\footnote{https://sourceforge.net/} etc. provide access to a large set of source code artifacts and their related information (e.g. the number of developers involved, number of bugs associated with it, people reporting the bugs etc.) \cite{allamanis2013mining}. Similarly, the bugs reported corresponding to various OSS software are usually tracked and available through different bug tracking portals \cite{ye2014learning} of major OSS foundations such as Apache\footnote{https://bz.apache.org/bugzilla/} and Eclipse\footnote{https://bugs.eclipse.org/bugs/} etc. Availability of such a rich set of raw data offers a ripe opportunity to extract and exploit the latent knowledge present in it. 

One possible direction for extracting and exploiting such latent knowledge involves examining the \textit{programming style} features of the source code. Typically, a software engineer makes a variety of programming choices during the construction of a software application. For instance, some of the key choices include:
\begin{itemize}
\item Choosing a particular programming language (e.g. Java over C\#).
\item Choosing a particular programming construct (e.g. \texttt{if} statement over the \texttt{switch} statement) to be used in a scenario.
\item Choosing a particular form of identifier name (e.g. a long but descriptive identifier name or a small and an ambiguous one \cite{lawrie2006s}).
\item Program design decisions \cite{allamanis2014learning}.
\end{itemize}
All such choices that a programmer makes define the \textit{programming style} of a programmer. The programming style used in a software influences several quality attributes of the software, such as readability, portability, ease-of-learning \cite{malheiros2012source}, reliability and maintainability \cite{allamanis2014learning}. 
For instance, the choice of concise and consistent naming conventions for identifier names has been reported to result in a better quality software \cite{deissenboeck2006concise}. 

\begin{figure}
\includegraphics[height=60mm, width=80mm]{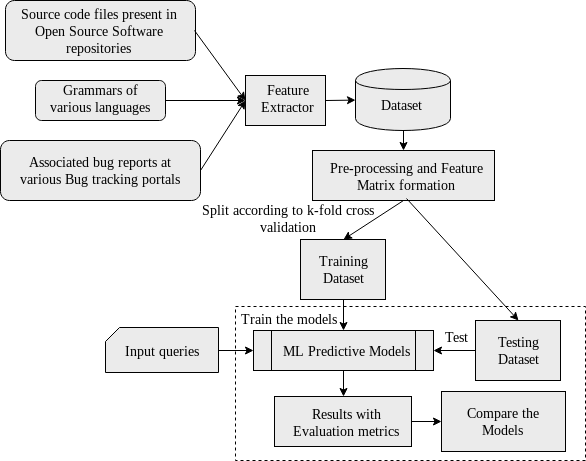}
 \setlength{\belowcaptionskip}{-15pt}
\caption{Predictive Model}
\label{fig:predictive-model}
\end{figure}

In our approach we have exploited the relationship that exists between programming style and defects associated with a source code. Basic idea of our approach is as follows (also depicted in Figure \ref{fig:basic-approach}):
\begin{itemize}
\item We first extract the programming style information (\textit{shallow knowledge}) associated with source files taken from various OSS repositories. 
The \textit{shallow knowledge} is stored in a relational database.
\item
\textit{Refined Knowledge} is then created by first extracting a subset of the shallow knowledge (created in the previous set) and then applying a couple of transformations on this knowledge to avoid dataset biasness as discussed in Section \ref{sec:dataset-creation}. 

\item
Using the dataset created, we then build a Machine Learning (ML) model, as shown in Figure \ref{fig:predictive-model}, which estimates the defectiveness associated with an input source file. An exhaustive comparison of the results obtained corresponding to various ML techniques used for prediction is presented in Figure \ref{fig:phase1-results}-\ref{fig:priority-results}. 
\end{itemize}
The software artifacts thus created, can be further used to build tools and techniques pertaining to several automated software engineering areas like bug localization \cite{zhou2012should, hindle2012naturalness}, code review and recommendation \cite{malheiros2012source, hindle2012naturalness} and program repair \cite{kim2013automatic} etc. The rest of the paper is organized as follows. Section \ref{sec:solution} describes the details of the proposed solution. We present the details about our implementation in Section \ref{sec:implementation}, and in Section \ref{sec:results} we discuss the results and analysis of our experiments. Section \ref{sec:related-work} describes the related work. Conclusions drawn from our work are presented in Section \ref{sec:conclusion}.

\section{Proposed solution}
\label{sec:solution}

\begin{figure}
\includegraphics[height=60mm, width=100mm]{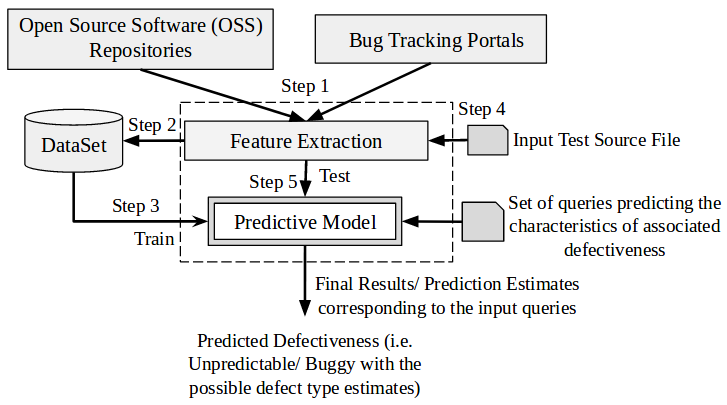}
  \setlength{\abovecaptionskip}{-5pt}
 \setlength{\belowcaptionskip}{-15pt}
\caption{Outline of the proposed approach}
\label{fig:basic-approach}
\end{figure}

A broad outline of the proposed solution is shown in Figure \ref{fig:basic-approach}. The \textit{Predictive Model}, shown as the central entity in Figure \ref{fig:basic-approach}, estimates defectiveness associated with an input source file. A predictive model typically is trained using pertinent information about the intended inputs. Thus, for our case a \textit{dataset} is built by extracting useful knowledge from  source code files and their corresponding defect reports. We achieve this by extracting information about programming styles used and the corresponding bugs, if any reported, at various OSS repositories and bug tracking portals. Thus the entire work can be split into two major tasks: 1) Fetching raw data and creation of a structured dataset, 2) Building the predictive model to estimate the defectiveness for given input source code.

\subsection{Creating a structured dataset}
\label{sec:dataset-creation}
\begin{figure}
\includegraphics[height=70mm, width=85mm]{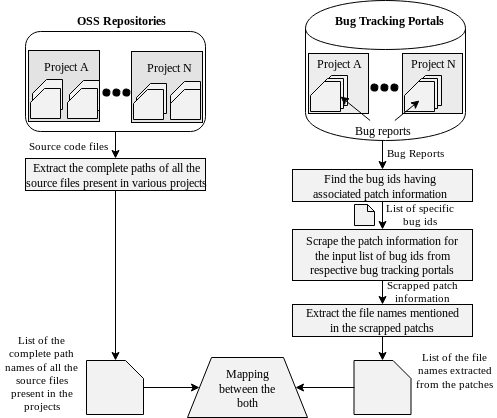}
  \setlength{\abovecaptionskip}{-5pt}
 \setlength{\belowcaptionskip}{-15pt}
\caption{Establishing mapping between the source files and bug reports}
\label{fig:mapping}
\end{figure}

\begin{figure}
\includegraphics[height=90mm, width=90mm]{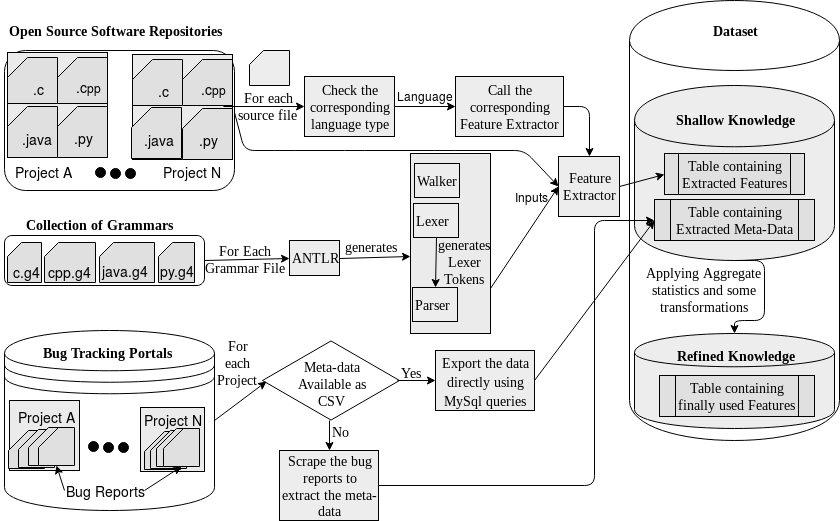}
\setlength{\belowcaptionskip}{-5pt}
\caption{Dataset Creation}
\label{fig:dataset}
\end{figure}

It has been shown \cite{deissenboeck2006concise,allamanis2014learning} that programming style significantly influences the quality of software. Based on these results we decided to make use of programming style information as one of the primary feature for training our prediction models (discussed shortly). Thus, we extracted programming style information by processing a corpus of source code files written in different programming languages. We also extracted relevant information from available bug reports corresponding to the source code files that we considered. Following are the main steps involved:
\begin{enumerate}
\item \textit{Identify suitable source code repositories and bug tracking portals:} The code repositories such as GitHub, SourceForge etc. contain large number of open source projects. Similarly, the bug tracking portals for major OSS foundations such as Apache and Eclipse etc. track bugs corresponding to such OSS projects. Several OSS projects on GitHub have sufficient source files having at least one bug reported against it. Therefore we took bulk of the raw content (source files) from GitHub for building our dataset.

\item \textit{Identify suitable OSS repositories:} Java, Python, C++ and C seem to be the most used programming languages\footnote{Professional Developer Survey 2017 by StackOverflow: https://insights.stackoverflow.com/survey/2017}. Thus we considered mainly those OSS repositories that have source code files written in one of these programming languages.

\item \textit{Establishing the mapping between the bug reports and the source files present in OSS repositories:} In order to find the source files associated with bug reports, we utilize the \emph{summary} and the \emph{patch} field of the bug reports. The \emph{patch} information associated with the bug reports mostly contains at least one mention of the associated source file. We use this information to establish a mapping between the source files present in the OSS repositories and the bugs reported corresponding to them as explained in Figure \ref{fig:mapping}.    

\item \textit{Feature extraction technique:} Raw source files need to be processed in order to extract relevant programming style attributes/features present in each. As such a custom lexer/ parser was built using ANTLR\footnote{http://www.antlr.org/} (Another Tool for Language Recognition). On providing a grammar file $G$ as an input, ANTLR generates a \textit{parse tree listener}\footnote{http://www.antlr.org/api/Java/org/antlr/v4/runtime/tree/ParseTreeListener.html}  class which contains methods for handling \textit{AST token}s as they are encountered during parsing of input source code conforming to the grammar $G$.

We overrode relevant methods of the parse tree listener for extracting desired features about the input source code. The features were computed using statistical measures such as average, min., max., standard deviation for values such as counts, lengths and depths of different programming constructs as enumerated in Table \ref{tab:metrics-details}.

\item \textit{Extract relevant un/semi- structured information (shallow knowledge):} 
The statistical measures captured in the previous step are aggregated at various levels (viz., function level, class level and file level) and stored in a database as \emph{shallow knowledge}. The detailed extraction process is depicted in Figure \ref{fig:dataset}. Also, for each source file which is referred to in a bug report we extract information such as \emph{priority, status, type and user exposure} from that bug report and store that information in the database.

\item \textit{Refine the shallow knowledge:} Refining the shallow knowledge is necessary for building a training dataset for a classifier.
Since the dataset used for training the ML models should not be biased by factors such as a particular language or file length or a particular feature we normalize the dataset. First step of normalization involves filtering only those source files (across programming languages) which are of \textit{similar size}. This was done for limiting the bias due to size of a source file. Further, the absolute values of various features extracted from a source file were divided by the file length. Similarly, the bias towards individual features was removed by using a MinMaxScalar\footnote{http://scikit-learn.org/stable/modules/generated/sklearn.preprocessing.MinMaxScaler.html} function of \texttt{Scikit\ -\ Learn\footnote{http://scikit-learn.org/}}.
The result of this step is our final dataset.
\end{enumerate}

\subsubsection{Composition of the dataset}
\label{sec:dataset}
\begin{table}
\small
\centering
  \caption{Language wise description of source files present in the dataset}
  \label{tab:language-wise-file-count}
 \resizebox{\columnwidth}{!}{
 \begin{tabular}{cccc}
  \hline
  \multirow{2}{*}{Language}&\multirow{2}{*}{Total file count}&\multirow{2}{*}{\shortstack{Total bug-linked\\ file count}}&\multirow{2}{*}{\shortstack{Candidate\\ file count\footnote{Files with the similar length chosen to build the feature matrix for training the ML models}}}\\
  &&&\\  
    \hline
    C&$117148$&$3224$&$173$\\
    C++&$7202$&$174$&$173$\\
    Java&$7500$&$318$&$173$\\
    Python&$4023$&$1440$&$173$\\
    \hline
\end{tabular}}
\end{table}
In order to eliminate the programming language selection bias we fetched the source code written in four different programming languages, viz., C, C++, Java and Python. A total of about $30400$ source files from $20$ different GitHub repositories were considered for creating the dataset. Further, about $14950$ bug reports associated with these source files were extracted from $4$ bug tracking portals of major OSS foundations, viz., Apache\footnote{https://bz.apache.org/bugzilla/}, GNU\footnote{https://gcc.gnu.org/bugzilla/}, Eclipse\footnote{https://bugs.eclipse.org/bugs} and  Python\footnote{https://bugs.python.org/}. Table \ref{tab:language-wise-file-count} and \ref{tab:repository-details} describe the composition of the dataset by providing the details such as the type of source files, OSS repositories and bug tracking portals chosen while building the dataset.

\begin{figure}
\includegraphics[height=70mm, width=85mm]{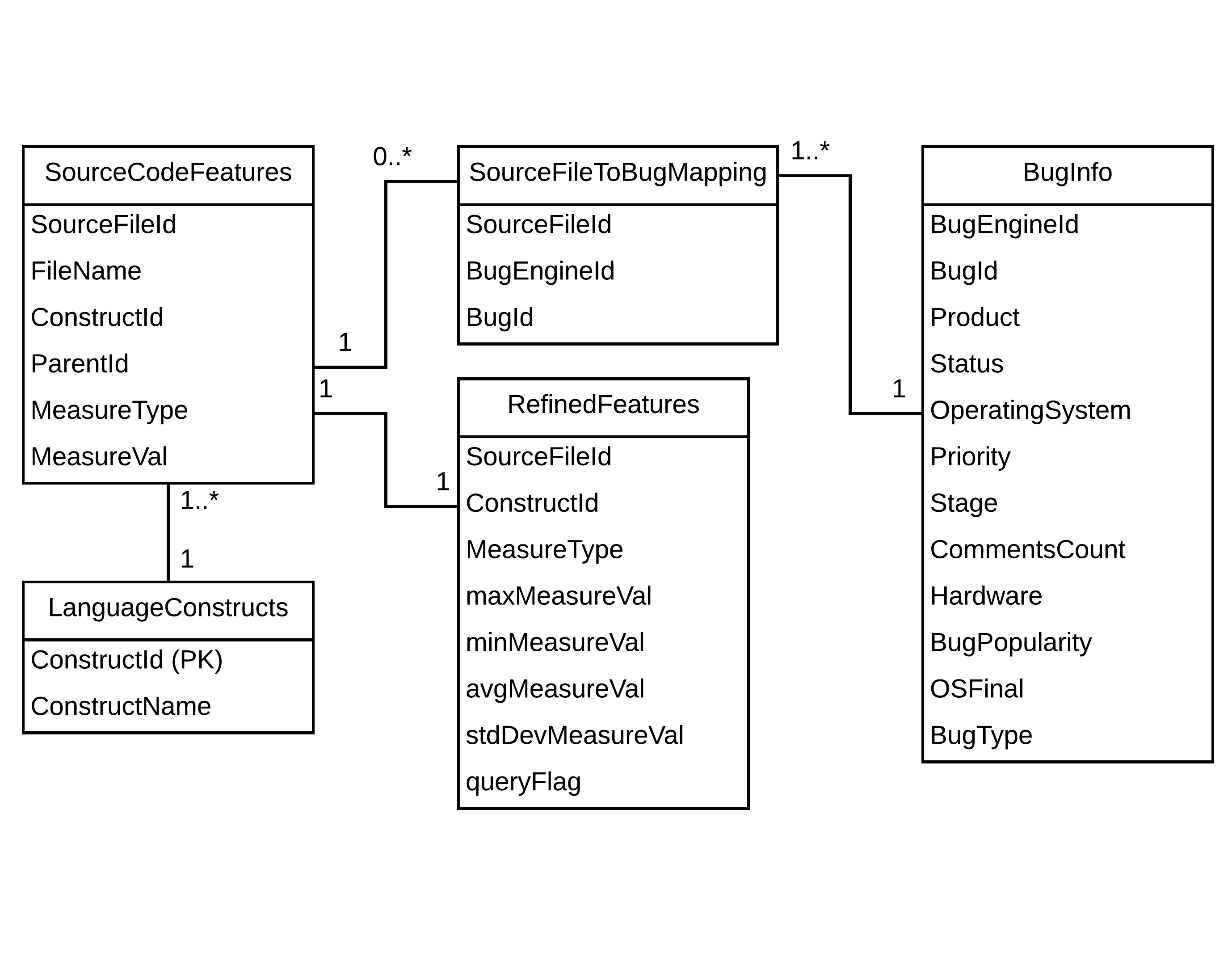}
  \setlength{\abovecaptionskip}{-25pt}
 \setlength{\belowcaptionskip}{-15pt}
\caption{Partial schema showing main entities of dataset.}
\label{fig:design}
\end{figure}

The information captured in our dataset is stored in a \textit{relational database}. Partial schema of the database showing main tables comprising our dataset are shown in Figure \ref{fig:design}. A brief description of these tables is as follows:
\begin{enumerate}
\item \texttt{SourceCodeFeatures}: It contains the features extracted (as described in Section \ref{sec:dataset-creation}) from various source files present in different OSS GitHub repositories.
\item \texttt{BugInfo}: It contains the relevant meta-data from associated bug reports to the source files considered in the previous \texttt{SourceCodeFeatures} table.
\item \texttt{SourceFileToBugMapping}: It stores the mapping between the source files and the bug reports considered in the dataset formation.
\item \texttt{LanguageConstructs}: It contains the information about the unique identifiers assigned to various language constructs. 
\item \texttt{RefinedFeatures}: It stores the refined features obtained by transforming the shallow knowledge as explained in Section \ref{sec:dataset-creation} step ($6$). 
\end{enumerate}

\begin{table}
\small
\centering
  \caption{Details of GitHub repositories and associated bug reports}
  \label{tab:repository-details}
  \resizebox{0.9\columnwidth}{!}{
  \begin{tabular}{c|c|c|c|c}
\toprule
  \multirow{3}{*}{\shortstack{OSS\\ repository}}&\multirow{3}{*}{\shortstack{Total\\ source\\ files}}&\multirow{3}{*}{\shortstack{Bug\\ tracking\\ portal}}&\multirow{3}{*}{\shortstack{Total\\ reported\\ bugs}}&\multirow{3}{*}{\shortstack{Total source\\ files linked\\ to bugs\footnote{Not every bug report provides information about the involved source file}}}\\
    &&&&\\
    &&&&\\
    \midrule[1pt]
    ant-master&$1220$&\multirow{12}{*}{Apache}&$653$&$96$\\
    \cline{1-2}
    \cline{4-5}
    batik-trunk&$1650$&&$251$&$104$\\
    \cline{1-2}
    \cline{4-5}
    \multirow{2}{*}{\shortstack{commons-bcel\\ -trunk}}&\multirow{2}{*}{$485$}&&\multirow{2}{*}{$29$}&\multirow{2}{*}{$2$}\\
     &&&&\\
     \cline{1-2}
    \cline{4-5}
    \multirow{2}{*}{\shortstack{lenya-BRANCH\\ -1-2-X}}&\multirow{2}{*}{$430$}&&\multirow{2}{*}{$234$}&\multirow{2}{*}{$16$}\\
    &&&&\\
    \cline{1-2}
    \cline{4-5}
    \multirow{2}{*}{\shortstack{webdavjedit\\ -master}}&\multirow{2}{*}{$7$}&&\multirow{2}{*}{$0$}&\multirow{2}{*}{$0$}\\
    &&&&\\
    \cline{1-2}
    \cline{4-5}
    poi-trunk&$3284$&&$103$&$86$\\
    \cline{1-2}
    \cline{4-5}
    \multirow{3}{*}{\shortstack{pengyou\\ -clients\\ -master}}&\multirow{3}{*}{$198$}&&\multirow{3}{*}{$15$}&\multirow{3}{*}{$11$}\\
    &&&&\\
    &&&&\\
    \hline
    gcc-master&$18524$&GCC&$6087$&$3296$\\
    \hline
    \multirow{3}{*}{\shortstack{org.eclipse.\\ -paho.mqtt.\\ -python-master}}&\multirow{3}{*}{$40$}& \multirow{8}{*}{Eclipse}&\multirow{3}{*}{$77$}&\multirow{3}{*}{$4$}\\
    &&&&\\
    &&&&\\
    \cline{1-2}
    \cline{4-5}
    \multirow{3}{*}{\shortstack{paho.mqtt.\\ -embedded\\ -c-master}}&\multirow{3}{*}{$33$}&&\multirow{3}{*}{$3$}&\multirow{3}{*}{$2$}\\
    &&&&\\
    &&&&\\
    \cline{1-2}
    \cline{4-5}
    \multirow{2}{*}{\shortstack{paho.mqtt.\\ -java-master}}&\multirow{2}{*}{$228$}&&\multirow{2}{*}{$5$}&\multirow{2}{*}{$3$}\\
    &&&&\\
 \hline
    cpython-master&$1336$& \multirow{10}{*}{Python}&$3235$&$525$\\
    \cline{1-2}
    \cline{4-5}
    bedevere-master&$16$&&$82$&$4$\\
    \cline{1-2}
    \cline{4-5}
    mypy-master&$182$&&$253$&$31$\\
    \cline{1-2}
    \cline{4-5}
    
    peps-master&$18$&&$20$&$4$\\
    \cline{1-2}
    \cline{4-5}
    planet-master&$16$&&$35$&$3$\\
    \cline{1-2}
    \cline{4-5}
    Python-2.7.14&$1325$&&$1521$&$443$\\
    \cline{1-2}
    \cline{4-5}
    Python-3.6.3&$1284$&&$1712$&$479$\\
    \cline{1-2}
    \cline{4-5}
    typeshed-master&$3$&&$0$&$0$\\
    \cline{1-2}
    \cline{4-5}
    \multirow{2}{*}{\shortstack{pythondotorg\\ -master}}&\multirow{2}{*}{$168$}&&\multirow{2}{*}{$642$}&\multirow{2}{*}{$47$}\\
    &&&&\\
    \hline
\end{tabular}
}
\end{table}

\begin{table}
\small
\centering
  \caption{Details of source code features}
  \label{tab:metrics-details}
 \resizebox{0.8\columnwidth}{!}{
 \begin{tabular}{c|c}
  \toprule
  Software Metric&Description\\
    \midrule[1pt]
    \texttt{\multirow{2}{*}{maxXCount}}&\multirow{2}{*}{\shortstack{maximum number of times an \texttt{X} construct\\ is used in a source code}}\\
   &\\
   \hline
    \texttt{\multirow{2}{*}{minXCount}}&\multirow{2}{*}{\shortstack{minimum number of times an \texttt{X} construct\\ is used in a source code}}\\
    &\\
    \hline
    \texttt{\multirow{2}{*}{avgXCount}}&\multirow{2}{*}{\shortstack{average number of times an \texttt{X} construct\\ is used in a source code}}\\
    &\\
    \hline
    \texttt{\multirow{2}{*}{stdDevXCount}}&\multirow{2}{*}{\shortstack{standard deviation of the number of times\\ an \texttt{X} construct is used in a source code}}\\
    &\\
    \hline
    \texttt{\multirow{2}{*}{maxXDepth}}&\multirow{2}{*}{\shortstack{maximum depth at which an \texttt{X}\\  construct is used in a source code}}\\
    &\\
    \hline
    \texttt{\multirow{2}{*}{minXDepth}}&\multirow{2}{*}{\shortstack{minimum depth at which an \texttt{X}\\ construct is used in a source code}}\\
    &\\
   \hline
   \texttt{\multirow{2}{*}{avgXDepth}}&\multirow{2}{*}{\shortstack{average depth at which an \texttt{X}\\  construct is used in a source code}}\\
    &\\
    \hline
    \texttt{\multirow{2}{*}{stdDevXDepth}}&\multirow{2}{*}{\shortstack{standard deviation of depth at which\\ an \texttt{X} construct is used in a source code}}\\
    &\\
   \hline
   \texttt{\multirow{2}{*}{maxXLength}}&\multirow{2}{*}{\shortstack{maximum length of an \texttt{X} construct\\ used in a source code}}\\
    &\\
   \hline
   \texttt{\multirow{2}{*}{minXLength}}&\multirow{2}{*}{\shortstack{minimum length of an \texttt{X} construct\\ used in a source code}}\\
    &\\
  \hline
  \texttt{\multirow{2}{*}{avgXLength}}&\multirow{2}{*}{\shortstack{average length of an \texttt{X}  construct\\ used in a source code}}\\
    &\\
\hline
\texttt{\multirow{2}{*}{stdDevXLength}}&\multirow{2}{*}{\shortstack{standard deviation of length of an\\  \texttt{X} construct used in a source code}}\\
    &\\
\hline
\end{tabular}}
\end{table}

A copy of the dataset that we used is available at \url{https://goo.gl/wqzzLU}.
\subsection{Building the predictive model}
\label{sec:predictive-model}

Next important task in our system is to estimate the defectiveness of an input source code file. For this we have to identify the most accurate ML algorithm with its parameters tuned for the problem. We make use of the dataset described in Section \ref{sec:dataset} to train $49$ different ML models (henceforth referred as \textit{Predictive Model}) as indicated in Table \ref{tab:ml-models}. We then select the best performing model based on standard accuracy metric scores (discussed in Section \ref{sec:results}).

The Predictive Model estimates the defectiveness of an input source code file in two phases: a) it first classifies the file as \texttt{likely- to-be-defective} or \texttt{unpredictable}, and b) if the input source file is classified as \texttt{likely-to-be-defective} then we label the file with prominent characteristics of bugs (for instance, \texttt{type} and \texttt{severity} of the bugs etc.). Basically, the second phase deals with determining the characteristics of likely bugs. It seeks to provide the answers to questions such as:  What is the probability that the input file is remarked as \texttt{likely-to-be-defective} contains bugs:
\begin{enumerate}
\item of a specific priority/severity (e.g. \texttt{Critical, High, Low, Medium} etc.)?
\item of a specific type (e.g. \texttt{Enhancement, BugFix} etc.)?
\item that are manifesting on a specific operating system (OS)?
\item that are manifesting on a specific hardware?
\item that involve a specific level of user exposure (measured, for instance, via \texttt{number of comments} on the bug reports)?
\end{enumerate}

The corresponding results are shown in Figure \ref{fig:phase1-results}-\ref{fig:priority-results} and discussed in Section \ref{sec:results}. Although, we have performed the experiments on a subset of bug characteristics, the work can be repeated for the other bug characteristics too. Further, it is not difficult to repeat (we plan to do it in future) the experiments for predicting defectiveness along various qualitative scenarios such as:
\begin{itemize}
\item Which programming language is likely to lead to defective programs?
\item Which type of bugs (enhancements, defects etc.) are most likely on a particular OS/hardware/programming language etc.
\end{itemize}

\section{Implementation details}
\label{sec:implementation}  
\begin{table}
\small
\centering
\caption{ML. models used in different phases of the predictive model}
\label{tab:ml-models}
\tabcolsep=0.11cm
\resizebox{0.8\linewidth}{!}{
 \begin{tabular}{c|c|c}
  \toprule
  Key&Phase 1&Phase 2\\
    \midrule[1pt]
    
    a&\multicolumn{2}{c}{LSVM($0.1$, \lq{ovr}\rq)}\\
    \hline
    b&\multicolumn{2}{c}{LSVM($0.1$, \lq{ovo}\rq)}\\
    \hline
    c&\multicolumn{2}{c}{LSVM($1.0$, \lq{ovr}\rq)}\\
    \hline
    d&\multicolumn{2}{c}{LSVM($1.0$, \lq{ovo}\rq)}\\
    \hline
    e&SVM($1.0$, \lq{ovr}\rq, \lq{l}\rq)&LSVM($0.5$, \lq{ovr}\rq)\\
    \hline
    f&SVM($1.0$, \lq{ovo}\rq, \lq{l}\rq)&LSVM($0.5$, \lq{ovo}\rq)\\
    \hline
    g&SVM($10$, \lq{ovr}\rq, \lq{r}\rq, $0.2$)&LSM($5$, \lq{ovr}\rq)\\
    \hline
    h&SVM($10$, \lq{ovo}\rq, \lq{r}\rq, $0.2$)&LSM($5$, \lq{ovo}\rq)\\
    \hline
    i&SVM($1.0$, \lq{ovo}\rq, \lq{p}\rq, $2$)&LSM($10$, \lq{ovo}\rq)\\
    \hline
    j&SVM($1.0$, \lq{ovr}\rq, \lq{p}\rq, $2$)&LSM($10$, \lq{ovo}\rq)\\
    \hline
    k&SVM($1.0$, \lq{ovo}\rq, \lq{p}\rq, $3$)&SVM($1.0$, \lq{ovr}\rq, \lq{l}\rq)\\
    \hline
    l&SVM($1.0$, \lq{ovr}\rq, \lq{p}\rq, $3$)&SVM($1.0$, \lq{ovo}\rq, \lq{l}\rq)\\
    \hline
    m&SVM($1.0$, \lq{ovr}\rq, \lq{s}\rq, $3$)&SVM($10$, \lq{ovr}\rq, \lq{r}\rq, $0.2$)\\
    \hline
    n&SVM($1.0$, \lq{ovo}\rq, \lq{s}\rq, $3$)&SVM($10$, \lq{ovo}\rq, \lq{r}\rq, $0.2$)\\
    \hline
    o&RF($10$)&SVM($1.0$, \lq{ovo}\rq, \lq{p}\rq, $2$)\\
    \hline
    p&RF($5$)&SVM($1.0$, \lq{ovr}\rq, \lq{p}\rq, $2$)\\
    \hline
    q&RF($7$)&SVM($1.0$, \lq{ovo}\rq, \lq{p}\rq, $3$)\\
    \hline
    r&RF($20$)&SVM($1.0$, \lq{ovr}\rq, \lq{p}\rq, $3$)\\
    \hline
    s&RF($50$)&SVM($1.0$, \lq{ovr}\rq, \lq{s}\rq, $3$)\\
    \hline
    t&RF($100$)&SVM($1.0$, \lq{ovo}\rq, \lq{s}\rq, $3$)\\
    \hline
    u&NSVM($0.5$, \lq{l}\rq, \lq{ovo}\rq)&RF($10$)\\
    \hline
    v&NSVM($0.5$, \lq{l}\rq, \lq{ovr}\rq)&RF($7$)\\
    \hline
    w&NSVM($0.7$, \lq{l}\rq, \lq{ovo}\rq)&RF($20$)\\
    \hline
    x&NSVM($0.7$, \lq{l}\rq, \lq{ovr}\rq)&RF($50$)\\
    \hline
    y&NSVM($0.5$, \lq{r}\rq, \lq{ovo}\rq)&RF($75$)\\
    \hline
    z&NSVM($0.5$, \lq{r}\rq, \lq{ovr}\rq)&RF($100$)\\
    \hline
    A&NSVM($0.7$, \lq{r}\rq, \lq{ovo}\rq)&NSVM($0.5$, \lq{l}\rq, \lq{ovo}\rq)\\
    \hline
    B&NSVM($0.7$, \lq{r}\rq, \lq{ovr}\rq)&NSVM($0.5$, \lq{l}\rq, \lq{ovr}\rq)\\
    \hline
    C&NSVM($0.5$, \lq{r}\rq, \lq{ovo}\rq, $0.2$)&NSVM($0.1$, \lq{l}\rq, \lq{ovo}\rq)\\
    \hline
    D&NSVM($0.5$, \lq{r}\rq, \lq{ovr}\rq, $0.2$)&NSVM($0.1$, \lq{l}\rq, \lq{ovr}\rq)\\
    \hline
    E&NSVM($0.7$, \lq{r}\rq, \lq{ovo}\rq, $0.2$)&NSVM($0.5$, \lq{r}\rq, \lq{ovo}\rq, $0.2$)\\
    \hline
    F&NSVM($0.7$, \lq{r}\rq, \lq{ovr}\rq, $0.2$)&NSVM($0.5$, \lq{r}\rq, \lq{ovo}\rq, $0.2$)\\
    \hline
    G&NSVM($0.5$, \lq{s}\rq, \lq{ovo}\rq)&NSVM($0.5$, \lq{ovo}\rq, \lq{p}\rq, $3$)\\
    \hline
    H&NSVM($0.5$, \lq{s}\rq, \lq{ovr}\rq)&NSVM($0.5$, \lq{ovr}\rq, \lq{p}\rq, $3$)\\
    \hline
    I&NSVM($0.7$, \lq{s}\rq, \lq{ovo}\rq)&NSVM($0.1$, \lq{ovo}\rq, \lq{p}\rq, $3$)\\
    \hline
    J&NSVM($0.7$, \lq{s}\rq, \lq{ovr}\rq)&NSVM($0.1$, \lq{ovr}\rq, \lq{p}\rq, $3$)\\
    \hline
    K&NSVM($0.5$, \lq{p}\rq, \lq{ovo}\rq)&NSVM($0.5$, \lq{s}\rq, \lq{ovo}\rq)\\
    \hline
    L&NSVM($0.5$, \lq{p}\rq, \lq{ovr}\rq)&NSVM($0.5$, \lq{s}\rq, \lq{ovr}\rq)\\
    \hline
    M&NSVM($0.7$, \lq{p}\rq, \lq{ovr}\rq)&NSVM($0.7$, \lq{s}\rq, \lq{ovo}\rq)\\
    \hline
    N&NSVM($0.7$, \lq{p}\rq, \lq{ovo}\rq)&NSVM($0.7$, \lq{s}\rq, \lq{ovr}\rq)\\
    \hline
    O&NSVM($0.5$, \lq{p}\rq, \lq{ovo}\rq, $0.2$)&Gauss(\lq{r}\rq, \lq{ovo}\rq)\\
    \hline
    P&NSVM($0.5$, \lq{p}\rq, \lq{ovr}\rq, $0.2$)&Gauss(\lq{r}\rq, \lq{ovr}\rq)\\
    \hline
    Q&NSVM($0.7$, \lq{p}\rq, \lq{ovo}\rq, $0.2$)&KNN(\lq{e}\rq)\\
    \hline
    R&NSVM($0.7$, \lq{p}\rq, \lq{ovr}\rq, $0.2$)&KNN(\lq{m}\rq)\\
    \hline
    S&Gauss(\lq{r}\rq, \lq{ovo}\rq)&MLP\\
    \hline
    T&Gauss(\lq{r}\rq, \lq{ovr}\rq)&-\\
    \hline
    U&KNN&-\\
    \hline
    V&MLP&-\\
    \hline
\end{tabular}}
\end{table}
To obtain the best prediction results, training and testing for both phases is performed using 
variety of relevant 
$8$ ML classification techniques as described in Table \ref{tab:ml-models}. They include: \texttt{Linear SVM (LSVM)\cite{fan2008liblinear}, SVM and Nu-SVM (NSVM) with radial, sigmoid and poly kernels (based on libSVM \cite{chang2011libsvm}), Gaussian Process (Gauss) classifier\cite{rasmussen2004gaussian}, K Nearest Neighbors (KNN) \-   classifier\cite{dasarathy1991nearest}, Random Forest (RF) classifier\cite{breiman2001random} and Multi-Layer Perceptron (MLP)\cite{hinton1990connectionist} classifier}. We have used the implementation of these algorithms provided in \texttt{Scikit\ -\ Learn\footnote{http://scikit-learn.org/}}. Tuning of algorithm specific parameters was performed by carrying out several experiments by using $49$ different parameter configurations of these ML techniques. A brief description of the pertinent parameters (as mentioned in Table \ref{tab:ml-models}) of different ML algorithms that we tuned is as follows:

For \texttt{LSVM}, among the two input parameters i.e. for \texttt{LSVM(a,b)}, \texttt{a} refers to the \texttt{penalty} parameter and \texttt{b=\lq{ovr}\rq} represents to the method of classification (where \texttt{b=\lq{ovr}\rq} refers to the \texttt{one-vs-rest} approach and \lq{\texttt{ovo}}\rq refers to the \texttt{one-vs-one} approach); For \texttt{SVM}, the first two input parameters remain the same while the third (say \texttt{c}) represents the \texttt{kernel} parameter (where \texttt{c=\lq{l}\rq, \lq{r}\rq, \lq{p}\rq and \lq{s}\rq} refer to a \texttt{linear, radial, polynomial} and \texttt{sigmoid} kernel respectively) and the fourth represents the degree in case of a polynomial kernel (\texttt{\lq{p}\rq}); \texttt{RF} has only one input parameter which specifies the \texttt{number of decision trees} or the \texttt{number of estimators}; \texttt{NSVM} has the first parameter as \texttt{nu}, second as \texttt{kernel}, third as the \texttt{method of classification} and fourth as \texttt{gamma}. For \texttt{KNN}, the \texttt{\lq{e}\rq} input parameter represents the use of \texttt{euclidean distance} whereas the \texttt{\lq{m}\rq} input parameter represents the use of \texttt{Manhattan distance}. 

\subsection{Evaluating effectiveness of the system}
\label{sec:evaluating-effectiveness}
To measure the efficacy of our system we compute standard \cite{michalski2013machine} evaluation metrics of \texttt{Precision, Recall} and \texttt{F1 score} for the prediction models. Literature \cite{kim2013should}, \cite{wang2016automatically} and \cite{williams2005automatic} represents the use of these metrics to compute the performance effectiveness of similar predictive systems.  The respective equations are: 
\begin{align}
Precision = \frac{true\ positive}{true\ positive + false\ positive}
\end{align}
\begin{align}
Recall = \frac{true\ positive}{true\ positive + false\ negative}
\end{align}
\begin{align}
F1\ score = \frac{2*Precision*Recall}{Precision + Recall}
\end{align}

Since \texttt{F1 score} captures both the effect of \texttt{Precision} and \texttt{Recall}, we show the results corresponding to \texttt{F1 score} values and its respective \texttt{standard deviation values (or error values)} only. Higher is the \texttt{F1 score} better is the \texttt{prediction accuracy} of the model. Further, all the results obtained are validated using \texttt{k-fold cross validation (with k=5)}. Thus the evaluation metrics illustrated are the averaged values over all \texttt{k folds}. Results of the experiments are discussed in Section \ref{sec:results}.

\section{Results and analysis}
\label{sec:results}
A major goal of our experiments was to assess the efficacy of the proposed system. The system can be considered effective if it accurately estimates the defectiveness of input source code. This required us to identify the best ML model which could accurately estimate the defectiveness of an input source code. As such we present and discuss here only the accuracy metrics (please see Section \ref{sec:evaluating-effectiveness}) achieved by using different ML algorithms on our dataset, when answering prediction questions involving different scenarios. For instance, our aim in the experiments has been to identify the best performing algorithm for: 
\begin{itemize}
\item The \texttt{likely-to-be-defective, unpredictable} classification task on the input source code.
\item Estimating associated bug characteristics for a \texttt{likely-to-\ be-defective} labeled input. 
\end{itemize}

Salient observations from our experiments are discussed next.

\subsection{Phase-I prediction results}
\label{sec:phase1-results}
\begin{figure*}
\includegraphics[height=30mm, width=0.8\textwidth]{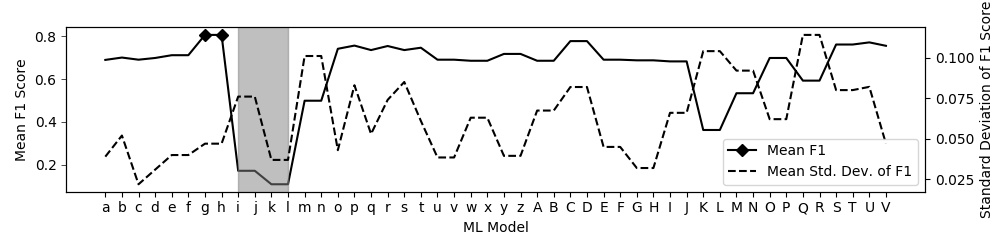}
 \setlength{\abovecaptionskip}{-0.5pt}
 \setlength{\belowcaptionskip}{-10pt}
\caption{Phase 1 results}
\label{fig:phase1-results}
\end{figure*}

Phase-I performs the task of classifying an input source file as \texttt{likely-to-be-defective} or \texttt{Unpredictable}. Accuracy metrics achieved by different ML models when testing on our dataset are shown in Figure \ref{fig:phase1-results}. Salient observations are as follows: \textbf{a)} \texttt{SVM with radial kernel} gives the highest (best) \texttt{averaged F1 score} of $0.807$ with a \texttt{standard deviation ($\sigma$)} of $0.047$. \textbf{b)} \texttt{SVM classifier with a poly kernel} gives the lowest (worst) \texttt{F1 score} on the dataset. The shaded portion in Figure \ref{fig:phase1-results} shows points corresponding to the lowest \texttt{F1 score} while the points marked with dark diamonds represent the models that yield the highest \texttt{F1 score}.

These accuracy metrics are obtained for the best of respective ML algorithm configurations. The tuning parameter values used for different scenarios are depicted in Table \ref{tab:ml-models}.

\begin{figure*}
\includegraphics[height=30mm, width=0.8\textwidth]{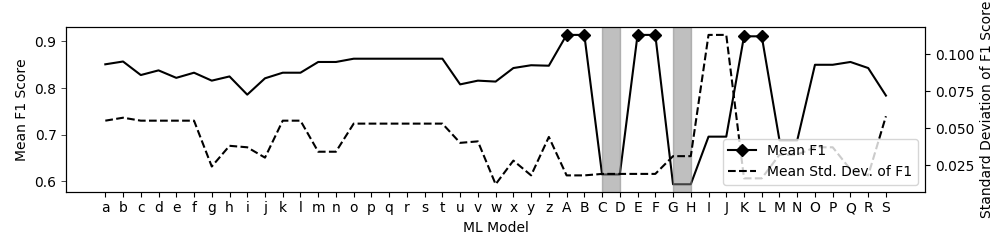}
\setlength{\abovecaptionskip}{-0.5pt}
 \setlength{\belowcaptionskip}{-10pt}
\caption{Prediction Results for bugs with \texttt{medium user exposure}}
\label{fig:popularity-results}
\end{figure*}
\begin{figure*}
\includegraphics[height=30mm, width=0.8\textwidth]{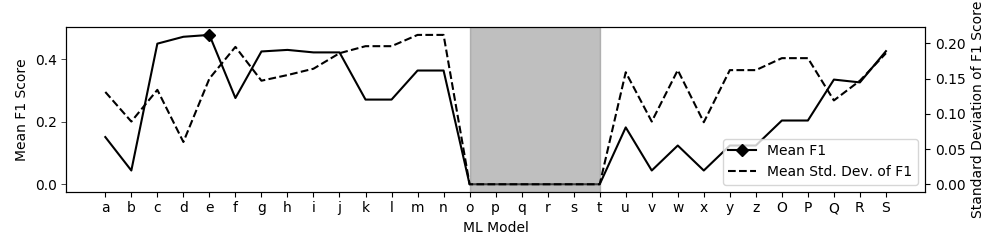}
\setlength{\abovecaptionskip}{-0.5pt}
 \setlength{\belowcaptionskip}{-10pt}
\caption{Prediction Results for bugs associated with \texttt{Windows} operating system}
\label{fig:os-results}
\end{figure*}
\begin{figure*}
\includegraphics[height=30mm, width=0.8\textwidth]{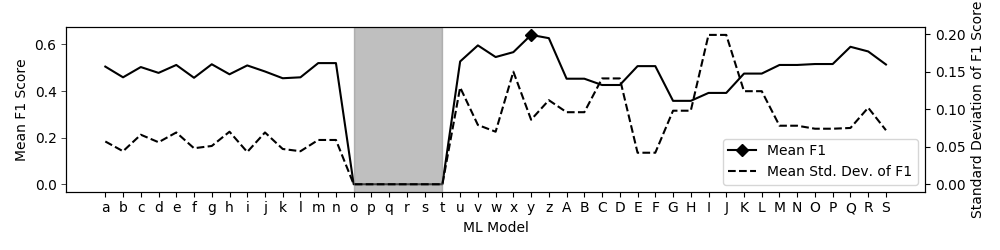}
\setlength{\abovecaptionskip}{-0.5pt}
 \setlength{\belowcaptionskip}{-10pt}
\caption{Prediction Results for bugs of type \texttt{Enhancement}}
\label{fig:type-results}
\end{figure*}
\begin{figure*}
\includegraphics[height=30mm, width=0.8\textwidth]{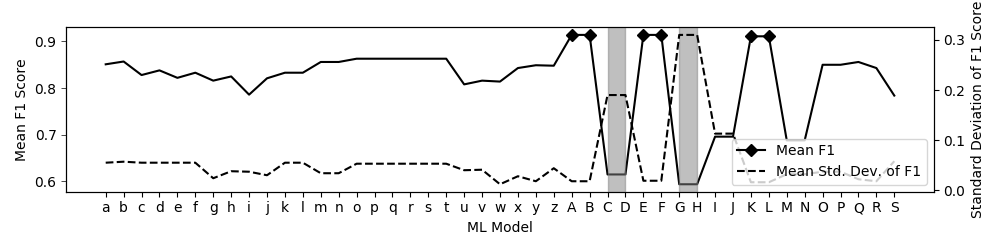}
\setlength{\abovecaptionskip}{-0.5pt}
 \setlength{\belowcaptionskip}{-10pt}
\caption{\small Prediction Results for bugs of \texttt{high priority}}
\label{fig:priority-results}
\end{figure*}

\subsection{Phase-II prediction results}
\label{sec:phase2-results}
Phase-II deals with predicting characteristics of the bugs that are likely to be associated with an input source code file that has been classified as \texttt{likely-to-be-defective} in the previous phase. 

Salient observations here are: \textbf{a)} For predicting \texttt{Medium exposure} bugs \texttt{Nu-SVM model} outperforms the rest -- it achieves an \texttt{F1 score} of $0.914$ and an associated $\sigma$ of $0.018$. \textbf{b)} \texttt{RF model with $75$ estimators (or decision trees)} yields the best \texttt{F1 score} ($0.641$ with $\sigma$ of $0.086$) for predicting bugs of type \texttt{Enhancement}. \textbf{c)} \texttt{RF model with $7$ estimators} achieves the best \texttt{F1 score} ($0.694$ and $\sigma$ of $0.036$) when predicting bugs annotated as \texttt{highest priority}. \textbf{d)} When predicting bugs annotated with a specific OS the \texttt{Linear SVM} gives the best \texttt{F1 score} ($0.478$ with $\sigma$ as $0.15$).

All these observations are shown in Figure \ref{fig:popularity-results}-\ref{fig:priority-results}. The shaded portion in these figures represents the points corresponding to the lowest \texttt{F1 score}. In most cases the shaded portion corresponds to the \texttt{SVM classifier with a poly kernel} setup, implying that it fared the worst on the dataset used. Points marked with dark diamonds on the graphs correspond to the models that give highest \texttt{F1 score}.

\section{Related work}
\label{sec:related-work}
Effects of programming styles on the quality of software has been well studied and reported in research literature. Most such studies can be categorized into two broad areas -- Quality and maintenance of software. Authors in \cite{allamanis2014learning, allamanis2015suggesting, deissenboeck2006concise, lawrie2006s} have examined in-depth the use of various programming styles, e.g., use of descriptive identifier names for improving quality of the software.

One of the works that addressed problems similar to ours is \cite{wang2016automatically}. They propose a learning algorithm to automatically extract the semantic representation associated with a program. They train a Deep Belief Network (DBN) on the token vectors extracted from the Abstract Syntax Tree (AST) representations of various programs. They rely on trained DBNs to detect the differences in these token vectors extracted from an input to predict defects. 
Another related contribution is presented by \cite{kim2013should}. They propose a two phase recommendation model for suggesting the files to be fixed by analyzing the bug reports associated with them. The recommendation model in its first phase categorizes a file as \enquote{predictable} or \enquote{unpredictable} depending upon the adequacy of the available content of bug reports. Further, in its second phase, amongst the files categorized as predictable, it recommends the top k files to be fixed. They experimented on a limited set of projects, viz., \enquote{Firefox} and \enquote{Core} packages of Mozilla.  

An information retrieval technique based bug localization module \enquote{BugLocator} is proposed by \cite{zhou2012should}. The BugLocator uses the concept of textual similarity to find a set of bug reports similar to an input bug report, and using the linked source code tries to identify potential bugs related to input. Use of textual similarity for code identification can pose problems because often times a given programming task may be coded in more than one way. 

A source code recommender, Mentor, for the use of novice programmers is presented by \cite{malheiros2012source}. Their aim is to help avoiding unnecessary diversion of programmer's attention from the main tasks. Mentor is based on a Prediction by Pattern Matching (PPM) algorithm. The authors have compared the performance of PPM with LSI on three OSS projects (GTK+, GIMP and Hadoop) and claim PPM to be better. 
Similarly, \cite{hindle2012naturalness} applies N-gram language models to infer syntactic and semantic properties of a program so as to provide code completion facilities. 

Overall, some of the key gaps that we find in the majority of the existing works that address problems similar to ours can be summarized as follows.
\begin{itemize}
\item In works that use (or propose) source code feature extraction for defect prediction and localization, only a limited types of nodes from program's AST has been utilized. For instance, node types such as identifier nodes, operator nodes, scope nodes, user defined data types etc. have not been usually considered. In our work we have captured all such types of nodes as per a language's grammar.
\item While building the feature vectors, mere presence/absence of programming constructs is considered. In our work, however, we also capture additional characteristics implied/associated with the programming constructs. For example, \enquote{length}, \enquote{count} and \enquote{depth-of-occurrences} of various constructs.
\item Association of \enquote{programming styles} adopted in software with the characteristics of the defects reported against such source code has not been adequately studied in literature.
\item Last but not the least, most works reported their results using a very limited volume (less than $5$ projects) of source code and bug reports, if any. Our study spans more than $30400$ source code files written in $4$ different programming languages and taken from $20$ OSS repositories.
\end{itemize}

\section{Conclusions}
\label{sec:conclusion}
Software maintenance tasks consume significant \cite{bennett2000software, lientz1980software} resources during software development. It has been well established \cite{deissenboeck2006concise,allamanis2014learning} that the programming practices followed during construction of a software greatly impact the quality of software. Analysis of the latent features present in source code can thus offer a valuable avenue for building predictive systems for detecting potential defects in software. 

In this paper we have proposed a system which leverages the large volume of source code available in OSS projects and their associated bug reports to create a new dataset of program's features. We then use this dataset to train a variety of state-of-the-art ML models which can accurately estimate the defectiveness of a given source code. We have used $49$ setups of $8$ different ML algorithms to examine their prediction accuracy on our dataset. This allowed us to identify the best performing model for predicting defectiveness of source code under variety of scenarios. For instance, we have shown that \texttt{SVM with radial kernel} performs the best (\texttt{averaged F1 score} of $0.807$) for identifying a source file as potentially defective or not. Similarly, \texttt{RF model with $7$ estimators} performs best (\texttt{F1 score} $0.694$) when predicting bugs annotated with \texttt{highest priority}.


\bibliographystyle{ACM-Reference-Format}
\bibliography{defectiveness-estimation-bibliography} 

\end{document}